\documentclass[prd,showpacs,preprintnumbers,nofootinbib,twocolumn]{revtex4}%

\usepackage{amsmath,amssymb,amsfonts,amsthm}
\usepackage{bbm}
\usepackage{bm}
\usepackage{graphicx}
\usepackage[OT2,OT1]{fontenc}
\newcommand\cyr{%
\renewcommand\rmdefault{wncyr}%
\renewcommand\sfdefault{wncyss}%
\renewcommand\encodingdefault{OT2}%
\normalfont
\selectfont}
\DeclareTextFontCommand{\textcyr}{\cyr}

\newcommand{\be}{\begin{equation}}
\newcommand{\ee}{\end{equation}}
\newcommand{\ba}{\begin{eqnarray}}
\newcommand{\ea}{\end{eqnarray}}
\def\bs{\begin{subequations}}
\def\es{\end{subequations}}
\def\a{\alpha}
\def\b{\beta}
\def\g{\gamma}

\def\k{\kappa}

\def\Om{\Omega}

\def\G{\Gamma}

\def\s{\sigma}

\def\vp{\varphi}
\def\N{\nabla}

\def\cO{{\cal O}}

\def\cK{{\cal K}}

\def\p{\partial}
\def\d{\delta}
\def\B{\Box}
\def\tphi{\tilde\phi}
\newcommand{\Eq}[1]{(\ref{#1})}

\def\rme{e}
\def\rmd{d}
\def\rmi{i}

\begin{document}

\title{Nonlocal gravity and the diffusion equation}


\author{Gianluca Calcagni}
\email{calcagni@aei.mpg.de}
\affiliation{Max Planck Institute for Gravitational Physics (Albert Einstein Institute),
Am M\"uhlenberg 1, D-14476 Golm, Germany}

\author{Giuseppe Nardelli}
\email{nardelli@dmf.unicatt.it}
\affiliation{Dipartimento di Matematica e Fisica, Universit\`a Cattolica,
via Musei 41, 25121 Brescia, Italy\\
and INFN Gruppo Collegato di Trento, Universit\`a di Trento,
38100 Povo (Trento), Italy}

\date{April 28, 2010}

\begin{abstract}
We propose a nonlocal scalar-tensor model of gravity with pseudodifferential operators inspired by the effective action of $p$-adic string and string field theory on flat spacetime. An infinite number of derivatives act both on the metric and scalar field sector. The system is localized via the diffusion equation approach and its cosmology is studied. We find several exact dynamical solutions, also in the presence of a barotropic fluid, which are stationary in the diffusion flow. In particular, and contrary to standard general relativity, there exist solutions with exponential and power-law scale factor also in an open universe, as well as solutions with sudden future singularities or a bounce. Also, from the point of view of quantum field theory, spontaneous symmetry breaking can be naturally realized in the class of actions we consider.
\end{abstract}

\preprint{AEI-2010-053 \qquad\qquad\qquad arXiv:1004.5144}
\preprint{PHYSICAL REVIEW D {\bf 82}, 123518 (2010)}

\maketitle


\section{Introduction}

Many proposals for modified gravity have been invoked in the hope of finding new insights into the open issues of the standard cosmological model. Among them, theories with pseudodifferential operators have been favored with particular attention. A reason is that nonlocal theories can have very different ultraviolet properties with respect to ordinary second- or higher-order actions (including popular Gauss--Bonnet extensions) and, hence, could play a role near the big bang and as spacetime effective formulations of nonperturbative quantum gravity. String field theory (SFT) is a concrete realization of this notion where pseudodifferential operators of the form
\be\label{eb}
\rme^{r_*\B}
\ee
decorate the effective target action of the fields, where $r_*$ is a constant and $\B$ is the spacetime d'Alembertian. The imprint of nonlocal dynamics in the history of the early universe, or even as dark energy models, has motivated the study of cosmological models inspired by open SFT \cite{are04,AJ,AKV1,cutac,AK,AV,kos07,AJV,AV2,cuta2,Jou07,Jo081,ArK,Jo082,NuM,BMNR,KV,cuta6,Ver09}, the $p$-adic string \cite{Jo082,NuM,BBC,lid07,BC,cuta4,BK2}, or other nonlocal effective actions featuring the operators \Eq{eb} \cite{BMS,kho06} or inverse powers of the d'Alembertian \cite{SW,DW,NO,Jhi08,Koia,Koib,CENO,DeW,CENOZ}. When nonlocality is of the type \Eq{eb}, it can be conveniently manipulated with the diffusion equation approach, which has been developed and employed, in analytic and numerical fashion, under different formulations \cite{cuta2,Jou07,Jo081,Jo082,NuM,cuta6,cuta4,vol03,FGN,vla05,roll,cuta3,MuN,cuta5,cuta7}.

A rather common assumption in the literature of nonlocal fields in cosmology is that nonlocality is confined only to one sector of the model, while the others are \emph{local}. 
In the case of SFT-motivated actions, the nonlocal sector is matter (a scalar field) and gravity is local and with Einstein--Hilbert action:
\be\label{lnl}
S=S_{\rm nonloc}(\phi)+\frac{1}{2\k^2}\int \rmd^Dx\sqrt{-g}\,R\,,
\ee
where $D$ is the topological dimension of spacetime, $g$ is the determinant of the metric $g_{\mu\nu}$, $\mu=0,\dots,D-1$, $\k^2=8\pi G$ is Newton's constant, and $R$ is the Ricci curvature scalar. This ansatz has been dictated mainly by the urgency of understanding, in the broadest sense, (i) the dynamics of the yet-unclear nonlocal scalar field theories, (ii) the combined effect of curvature and nonlocality, and (iii) its possible consequences for phenomenology, in particular, in relation to cosmology (inflation, dark energy) and the modification of flat open SFT solutions (can cosmological friction damp the wild oscillations of the OSFT solution with marginal deformations? \cite{cutac,Jo082,BMNR,cuta6,FGN}).

Now that robust analytical and numerical methods have been established to solve nonlocal equations of motion, it would be highly desirable to address the conceptual inconsistency subjacent to Eq.~\Eq{lnl}. Not only would we like to define a model with nonlocality implemented in all sectors (and reproducing standard general relativity in the limit of weak nonlocality), but we want also to find nontrivial cosmological solutions. Such is the twofold objective of this paper.

The problem of nonlocal gravity can be faced under three independent perspectives, one motivated by string field theory, one purely phenomenological, and another a hybrid approach. In the first case, the Einstein--Hilbert action in Eq.~\Eq{lnl} is introduced by hand as an educated guess on ``how the effective SFT action of tachyon might look like in the presence of gravity.'' Gravity is minimally coupled with a tachyon-type or $p$-adic scalar field whose action is dictated or inspired by concrete Minkowski calculations. Obviously, a fully consistent effective tachyonic action should be derived from first principles in all its sectors. As far as gravity is concerned, the natural framework is closed SFT \cite{SZ,KKS,KuS,KS3,Zwi92,SZ1,SZ2,SZ3,OZ,YZ1,YZ2,Mic06,Moe1,Moe2,Moe3}, which features the same nonlocal operator \Eq{eb} of open SFT. The subject is rather intricate and, unfortunately, effective gravitational nonlocal actions are known only at the linear level \cite{KS3,Moe3}.\footnote{On the other hand, the \emph{local} low-energy effective field theory of the closed string tachyon-dilaton-graviton system is well understood also in its cosmological properties \cite{YZ3,Swa08}.}

Instead of facing the rigors of closed SFT some toy models have been considered, in particular the open-closed $p$-adic tachyonic action \cite{BF2,MoS,Vla06} and a closed SFT-inspired tachyon-tachyon model \cite{Ohm03}. However, the graviton is not included, thus leaving the (third) possibility to consider phenomenological actions where the matter sector is as close as possible to SFT or the $p$-adic string \cite{BCK}, while a nonlocal gravitational sector is built from reasonable requirements (mainly, that it contains the same type of pseudodifferential operators as the matter sector). Here we shall follow the third path.

Nonlocal gravity sectors have been constructed with inverse powers of the $\B$ operator \cite{SW,DW,NO,Jhi08,Koia,Koib,CENO,DeW,CENOZ} or more general kinetic functions \cite{kho06}, while keeping matter local. In \cite{BMS} a nonlocal total action has been proposed with nonminimal coupling between gravity and a scalar field, but the dynamical analysis therein does not go beyond cosmological solutions when the matter sector is switched off.

In Sec.~\ref{act} we adopt the diffusion equation method to infer the form of a solvable scalar-tensor nonlocal action with pseudodifferential operators of exponential type, Eq.~\Eq{eb}. This approach is chosen by virtue of its nonperturbative character, which does not require truncating the theory in order to find solutions, exact or asymptotic. Exact nonvacuum solutions of the equations of motion of a $p$-adic-like system will be found in Sec.~\ref{sols} for cosmological backgrounds; their classical stability is checked. The exact solutions are stationary along the diffusion flow (i.e., the diffusion equation is trivially satisfied) but the scalar and Hubble profiles as well as their dynamics are nontrivial. Notably, there exist solutions with exponential and power-law scale factor when intrinsic curvature is negative (Sec.~\ref{cos2}), as well as a most general class of explicit solutions for actions with conformal operators (Sec.~\ref{cos3}). We make some general remarks on nonstationary asymptotic solutions in Sec.~\ref{cos4}, showing that, for natural choices of the potential, the system realizes spontaneous symmetry breaking.

We are interested in a scalar-tensor theory where nonlocal effects are dominant, and such a system finds a natural application only in the early, inflationary, or bouncing universe (here we ignore applications of nonlocal models to dark energy). For this reason, in most of the paper we do not assume the presence of any cosmological fluid such as dust or radiation, but nevertheless inclusion of extra matter components is briefly discussed in Sec.~\ref{matte}. In Sec.~\ref{slac} the analysis is extended to another action with a kinetic operator similar to the one of the SFT tachyon; de Sitter and power-law exact solutions are found. Section \ref{disc} is devoted to discussion. In the Appendix we recall flat and curved exponential and power-law solutions in standard general relativity.


\section{Definition and dynamics}\label{act}


\subsection{$p$-adic-like action}\label{acti}

Let us begin for simplicity with the $p$-adic action in Minkowski spacetime,
\be\label{act0}
S=\int \rmd^Dx\left[\frac12 \phi \rme^{r_*\B}\phi-V(\phi)\right],
\ee
where $V$ is the field potential, $\phi$ has dimension $[\phi]=D/2$ in momentum units, and $r_*$ is a constant of dimension $[r_*]=-2$. Quite often one works in units where the characteristic energy scale $M$ of the system is absorbed (corresponding to $\a'=1$ units in string theory) and treats $r_*\sim M^{-2}$ as dimensionless. In the diffusion equation method \cite{cuta3,cuta7}, one introduces an auxiliary direction $r$ and a scalar field $\phi(r,x)$ which, for the moment, is not related to the solutions $\phi(x)$ of the original nonlocal action. By definition, this scalar diffuses according to the equation
\be\label{difeq1}
(\B+\p_r)\phi(r,x)=0\,,
\ee
where $x=x^\mu$ and $\B=\p_\mu\p^\mu$ is the usual d'Alembertian operator (later generalized to the curved case). Since
\be\label{comm}
\left[\B,\p_r\right]=0\,,
\ee
the action of the operators \Eq{eb} is simply a translation along $r$:
\be\label{trans}
\rme^{r_*\B}\phi(r,x)=\rme^{-r_*\p_r}\phi(r,x)=\phi(r-r_*,x)\,.
\ee
At this point the system \Eq{act0} can be solved by a well-established algorithm which, however, we will not need because we shall concentrate on ``stationary'' exact solutions. Anyway, we briefly summarize it here for the interested reader. The idea is to choose an appropriate initial (in $r$) field configuration $\phi(0,x)$ such that the scalar evaluated at the end of the diffusion flow at some $r=r_1$,
\be\label{1x}
\phi(r_1,x)=\phi(x)\,,
\ee
is indeed the scalar field in the action and a solution to the equations of motion. To check this, one looks at the spacetime asymptotic behaviour of the equations of motion and verifies if there exist certain values of the free parameter $r_1$ such that the profile \Eq{1x} is a solution. The existence of this solution is not always guaranteed because the diffusion equation does not encode information on the potential and, in most applications, the result is approximate. In practice, however, the level of the approximation [depending both on the chosen configuration $\phi(0,x)$ and potential $V$] is more than acceptable and the found profile \Eq{1x} is a genuine nonperturbative solution. In string theory this is perhaps not surprising \cite{cuta7}. As a consequence of Eq.~\Eq{trans}, the system becomes \emph{localized} in spacetime and its infinite number of degrees of freedom are transferred into the specification of the \emph{field} configuration $\phi(0,x)$. The resulting model has a well-defined restatement of the Cauchy problem \cite{cuta3} (this point, in diffusing and nondiffusing models with exponential operators as well as in other nonlocal models, has been discussed also in \cite{NuM,BK2,MuN,BK1}). Concrete examples are the exact solution for the $p$-adic string \cite{cuta4} and the approximate solutions for open string and superstring field theory \cite{roll,cuta5,cuta7}. 

When introducing a nontrivial metric $g_{\mu\nu}$, we have to include also a nonlocal action for it. As far as the diffusion method is concerned, so far this has not been done, and discussions have been limited to systems of the form \Eq{lnl} \cite{cuta2} or with more general local gravitational actions \cite{cuta6}. The reason has been mainly technical. Suppose decorating the gravitational action with the exponential operators \Eq{eb}. If we hope to solve the system analytically or semianalytically, it is reasonable to expect that a metric-derived field will obey the diffusion equation. For instance, this could be the metric itself $g_{\mu\nu}(r,x)$ or one of the Riemann invariants $R(g)\equiv R_{\mu\nu\cdots}(r,x)$; different choices are physically inequivalent. Whatever the choice, the fundamental commutation property \Eq{comm} is no longer valid, because now the d'Alembertian is coordinate dependent and the diffusion equation for the metric or Riemann invariant $R(g)$ is essentially \emph{nonlinear}:
\be
\N_\s(g)\N^\s(g)R(g)= -\p_r R(g)\,,
\ee
where $\N_\s$ is the covariant derivative. Accordingly,
\ba
\B^2 R(g) &=& -\B\p_r R(g)\neq -\p_r\B R(g)\nonumber\\
&\neq& \p_r^2R(g)\,,\nonumber\\
&\cdots&\nonumber\\
\B^nR(g) &\neq& (-1)^n\p_r^nR(g)\,, \nonumber
\ea
and the translation property \Eq{trans} breaks down for $R(g)$.

However, our main goal is not really to impose the diffusion equation in all sectors, but rather (i) to include an infinite number of derivatives of the metric field for consistency with the matter sector, and (ii) do it in while preserving the translation property wherever and whenever required. Therefore, we can look towards another direction: namely, giving up diffusion in metric fields,
\be
\p_r R(g)=0\,,
\ee
and modifying the diffusion equation with an extra curvature term. Here we take for simplicity a term depending only on the Ricci scalar:
\be\label{difeqR}
[\B+f(R)+\p_r]\phi=0\,,
\ee
where $f$ is a function of dimension $[f]=2$. Equation \Eq{act0} is modified as 
\be\label{act1}
S=\int d^Dx\sqrt{-g}\left[\frac12 \phi \rme^{r_*[\B+f(R)]}\phi-V(\phi)\right]\,.
\ee
The role of the $f(R)$ term is to realize nonlocality in the gravity sector: an infinite number of derivatives act on metric fields via the Baker--Campbell--Hausdorff formula. This way the same exponential operator acts on both sectors and one can still solve the equations of motion nonperturbatively and analytically, either exactly or approximately.

Equation \Eq{act1} is the total action for our scalar-tensor model. The scalar sector diffuses while the metric sector does not, but both sectors are nonlocal as desired. The resulting equations of motion can be localized [$\phi$ translates, $\rme^{r_*[\B+f(R)]}\phi(r,x)=\phi(r-r_*,x)$] and solved for a given fixed metric.

Equation \Eq{act1} has well-defined limits as a pure scalar field or gravitational theory.
Let $f(0)=0$. In the Minkowski limit, Eq.~\Eq{act1} reduces to Eq.~\Eq{act0}. On the other hand, when $\phi$ relaxes to a constant $\phi_0$ [local minimum of the potential, $V_0\equiv V(\phi_0)\equiv \phi_0^2\Lambda$] one has, up to an overall constant,
\be\label{gravo}
S\sim\int d^Dx\sqrt{-g}\left\{\rme^{r_*[\B+f(R)]}-2\Lambda\right\}\,,
\ee
which becomes an ordinary $f(R)$ theory in the small $r_*$ limit. In the same limit but keeping $\phi$ dynamical,
\be\label{lowe}
S\sim\int \rmd^Dx\sqrt{-g}\left\{\frac12 \phi [r_*\B+r_* f(R)]\phi-V(\phi)\right\}\,.
\ee
Later on we will find solutions of the full nonlocal action for the linear case $f(R)=\a R$, where $\a$ is a dimensionless constant which will be often set to a negative value. To recover the Einstein--Hilbert Lagrangian in Eq.~\Eq{lowe} when $\phi=1$, it will be sufficient to choose $r_*<0$. With this choice, the kinetic term for the scalar field has the wrong sign but this does not correspond to a ghost, since Eq.~\Eq{lowe} is only an effective scalar-tensor action. The full action \Eq{act1} is not plagued by any such instability at least in the scalar sector and at least on Minkowski background, because there are no physical perturbative poles for nontrivial nonlinear interactions (e.g., \cite{cuta4,BMS}), while there is a real pole for a quadratic potential.\footnote{For a general background, the equation of motion is $e^{r_*[\B+f(t,{\bf x})]}\phi-V'=0$, leading to a very complicated propagator in momentum space which cannot be written explicitly except for special backgrounds. However, for a nontrivial interaction (e.g., $\phi^n$, $n\geq3$) the Green equation $e^A \tilde G(k)=1$ does not have nontrivial zeroes for any operator $A$ and one has no perturbative excitations on any background and for any function $f$. When the eigenstates of $A$ are a complete basis for the functions $\tilde G(k)$, the statement can be proven explicitly case by case. For spacetimes with constant Ricci curvature or which are homogeneous, $f$ does not depend on spatial coordinates. Then, for a nontrivial interaction the propagator is $\tilde G(k^2)=e^{-r_*(f-k^2)}$ which, as announced, has no nontrivial zeroes. On the other hand, for a quadratic potential the propagator reads $\tilde G(k^2)=[e^{r_*(f-k^2)}-m^2]^{-1}
= -(2m^2)^{-1}+m^{-2} \sum_{n=-\infty}^{+\infty}[r_*(f-k^2)-\log m^2- 2\pi n \rmi]^{-1}
$, where we expanded \emph{\'a la} Mittag--Leffler. There are infinitely many perturbative states with complex-mass poles $-k^2=r_*^{-1}\log m^2-f+ 2\pi nr_*^{-1}\rmi$. The theory can be rendered stable by projecting out these modes according to certain mathematical prescriptions \cite{BK1,GKR}.} Notice also that, looking at the weak-nonlocality limit \Eq{lowe}, one has
\be\label{afor}
|\a|=\frac{1}{\k^2 r_*\phi_0^2}\sim\left(\frac{M_{\rm Pl}M}{\phi_0}\right)^2\,,
\ee
where $M_{\rm Pl}$ is the reduced Planck mass. Most of the stable solutions we shall find have $|\a|<1$,\footnote{The only exception is Eq.~\Eq{pls} for $p$ near 0 or $2/D$. However, solutions with these fine-tuned $p$'s do not inflate.} which might indicate a trans-Planckian problem similar to that in standard inflation, where the classical field acquires large values above the Planck scale. However, nonlocal effects, if they exist, should not be negligible in the very early universe and the approximation \Eq{lowe} might not be reliable is such regimes. Consequently, the problem stemming from Eq.~\Eq{afor} might be an artifact of the weak-nonlocality approximation.

Although the motivations underlying our ansatz \Eq{act1} differ from those of \cite{BMS}, we eventually got a very similar result. This is not completely unexpected because both the diffusion equation and the ultraviolet properties invoked in the construction of \cite{BMS} (absence of ghosts and asymptotic freedom) rely on Eq.~\Eq{eb}. The diffusion equation encodes some of the gauge symmetries of SFT at spacetime level \cite{cuta7} and, as such, gives rise to rather rigid physical properties.


\subsection{Equations of motion}\label{act2}

To find the equations of motion we need the variations
\ba
\delta \sqrt{-g}&=& -\frac12\,g_{\mu\nu}\,\sqrt{-g}\,\delta g^{\mu\nu}\,,\\
\delta R        &=& (R_{\mu\nu}+g_{\mu\nu}\,\B-\N_\mu\N_\nu)\,\delta g^{\mu\nu}\,,
\ea
where $\N_\nu V_\mu \equiv \p_\nu V_\mu-\G^\s_{\mu\nu}V_\s$ is the covariant derivative of a vector $V_\mu$ and the curved d'Alembertian on a scalar $\phi$ is 
\be\label{dal}
\Box\phi =\frac{1}{\sqrt{-g}}\p^\mu (\sqrt{-g}\p_\mu\phi)\,.
\ee
We will make use of the operator identity \cite{Yan02}
\be
\delta \rme^{r_* X}=\int_0^{r_*} \rmd s\, \rme^{sX}(\delta X)\rme^{(r_*-s)X}\,,
\ee
for a (differential) operator $X$.

Also, for two scalars $\Phi_1$ and $\Phi_2$ ($\to$ indicates integration by parts),
\ba
\sqrt{-g}\Phi_1\frac{\delta\B}{\delta g^{\mu\nu}} \Phi_2 &=& \Phi_1\p_\mu (\sqrt{-g}\p_\nu\Phi_2)\nonumber\\
&&-\frac12\sqrt{-g}\Phi_1(\p^\s g_{\mu\nu})\p_\s\Phi_2\nonumber\\
&\to&\sqrt{-g}\left[\frac12 g_{\mu\nu} (\Phi_1\B \Phi_2+\p_\s \Phi_1 \p^\s \Phi_2)\right.\nonumber\\
&&\left.-(\p_\mu \Phi_1)( \p_\nu \Phi_2)\right]\,,
\ea
and
\ba
\sqrt{-g}\Phi_1\Phi_2\frac{\delta f(R)}{\delta g^{\mu\nu}}&=&\sqrt{-g}\Phi_1\Phi_2f'(R)\frac{\delta R}{\delta g^{\mu\nu}}\nonumber\\
&\to& \sqrt{-g}\left\{\Phi_1\Phi_2 f'(R)R_{\mu\nu}\right.\nonumber\\
&&\left.+(g_{\mu\nu}\B-\N_\mu\N_\nu)[\Phi_1\Phi_2f'(R)]\right\},\nonumber\\
\ea
where we have discarded boundary terms (this can be done compatibly with the variational principle after adding a boundary piece to the total action).

The scalar equation of motion $\delta S/\delta\phi=0$ is
\be\label{sceq}
\phi(1-r_*,x)=V'[\phi(1,x)]\,.
\ee
In the absence of extra matter, the Einstein equations are (coordinate dependence of the fields implicit)
\ba
0&=& \frac{2}{\sqrt{-g}}\frac{\delta S}{\delta g^{\mu\nu}}\nonumber\\
 &=& -\,g_{\mu\nu}\left\{\frac12 \phi(1)\phi(1-r_*)-V[\phi(1)]\right\}\nonumber\\
 &&+\frac{1}{\sqrt{-g}}\int_0^{r_*} \rmd s\,\int \rmd^Dx\sqrt{-g}\,\phi(1-s)\nonumber\\
 &&\times\left[\frac{\delta \B}{\delta g^{\mu\nu}}+\frac{\delta f(R)}{\delta g^{\mu\nu}}\right]\phi(1-r_*+s)\nonumber\\
 &=& -g_{\mu\nu}\left\{\frac12 \phi(1)\phi(1-r_*)-V[\phi(1)]\right\}\nonumber\\
 &&+\int_0^{r_*} \rmd s\,\Sigma_{\mu\nu}(s)\,,\label{EE1}\\
\Sigma_{\mu\nu}(s)&=& \frac12 g_{\mu\nu} [\phi(1-s)\B\phi(1-r_*+s)\nonumber\\
 &&+\p_\s \phi(1-s) \p^\s\phi(1-r_*+s)]\nonumber\\
&&-\p_\mu \phi(1-s)\p_\nu\phi(1-r_*+s)\nonumber\\
 &&+\phi(1-s)\phi(1-r_*+s) f'(R)R_{\mu\nu}\nonumber\\
&&+(g_{\mu\nu}\B-\N_\mu\N_\nu)[\phi(1-s)\nonumber\\
 &&\times\phi(1-r_*+s)f'(R)]\,.\label{EE2}
\ea
Taking the trace,
\ba
0&=&-D\left\{\frac12 \phi(1)\phi(1-r_*)-V[\phi(1)]\right\}+\int_0^{r_*} \rmd s\,\Sigma(s)\,,\nonumber\\&&\label{tr0}\\
\Sigma(s)&=& \frac{D}2\phi(1-s)\B\phi(1-r_*+s)\nonumber\\
 &&+\left(\frac{D}2-1\right)\p_\s \phi(1-s) \p^\s\phi(1-r_*+s)\nonumber\\
&&+\phi(1-s)\phi(1-r_*+s) f'(R)R\nonumber\\
 &&+(D-1)\B[\phi(1-s)\phi(1-r_*+s)f'(R)]\,.\label{tr}
\ea


\section{Cosmological solutions}\label{sols} 

A natural background whereon to study the model is Friedmann--Robertson--Walker (FRW), defined by the line element
\be
\rmd s^2=g_{\mu\nu}\rmd x^\mu\rmd x^\nu=-\rmd t^2+a(t)^2\tilde g_{ij}\rmd x^i\rmd x^j\,,
\ee
where $t$ is synchronous time, $a(t)$ is the scale factor, and
\be
\tilde g_{ij}\rmd x^i \rmd x^j=  \frac{\rmd \varrho^2}{1-\textsc{k}\,\varrho^2}+\varrho^2\rmd\Om^2_{D-2}
\ee
is the line element of the maximally symmetric $(D-1)$-dimensional space $\tilde\Sigma$ of constant sectional curvature $\textsc{k}$ (equal to $-1$ for an open universe, 0 for a flat universe, and $+1$ for a closed universe with radius $a$).


\subsection{Friedmann equations}\label{cos1}

On an FRW background,
\ba
&& R_{00}   = -(D-1)(H^2+\dot H)\,,\\
&& R_{i\!j} =\tilde R g_{i\!j}\equiv\left[\frac{2\textsc{k}}{a^2}+(D-1)H^2+\dot H\right]g_{i\!j}\,,\\
&& R = (D-1)\left(\frac{2\textsc{k}}{a^2}+DH^2+2\dot{H}\right),
\ea
where 
\be
H\equiv \frac{\dot{a}}{a}
\ee
is the Hubble parameter. The 00 component of Einstein's equations is
\ba
0&=& \frac12 \phi(1)\phi(1-r_*)-V[\phi(1)]+\int_0^{r_*} \rmd s\,\Sigma_{00}(s)\,,\nonumber\\
 &&\label{fr0}\\
\Sigma_{00}(s)&=& -\frac12\phi(1-s)\B\phi(1-r_*+s)\nonumber\\
 &&-\frac12\dot\phi(1-s)\dot\phi(1-r_*+s)\nonumber\\
&&+\phi(1-s)\phi(1-r_*+s) f'(R)R_{00}\nonumber\\
&&+(D-1)H\p_t[\phi(1-s)\phi(1-r_*+s)f'(R)].\nonumber\\
\ea
We specialize to the case
\be
f(R)=\a R\,,
\ee
where $\a$ is a dimensionless constant. In the Friedmann equations \Eq{fr0} and \Eq{tr0} one has
\ba
\Sigma_{00}(s)&=& -\frac12\phi(1-s)\B\phi(1-r_*+s)\nonumber\\
&&-\frac12\dot\phi(1-s)\dot\phi(1-r_*+s)\nonumber\\
&&-\a(D-1)(H^2+\dot H)\phi(1-s)\phi(1-r_*+s)\nonumber\\
&&+\a(D-1)H\p_t[\phi(1-s)\phi(1-r_*+s)]\,,
\ea
and
\ba
\Sigma(s)&=& \frac{D}2\phi(1-s)\B\phi(1-r_*+s)\nonumber\\
&&-\left(\frac{D}2-1\right)\dot\phi(1-s) \dot\phi(1-r_*+s)\nonumber\\
&&+\a\phi(1-s)\phi(1-r_*+s)R\nonumber\\
&&+\a(D-1)\B[\phi(1-s)\phi(1-r_*+s)].
\ea


\subsection{``Stationary'' solutions with quadratic potential}\label{cos2}

Although the diffusion method allows one to find asymptotic solutions for any given background, it is not necessary to resort to all its machinery in order to get some interesting results. In particular, one can look for simple solutions which are trivial along the diffusion flow, i.e., 
\be
\phi(r,t)=\rme^{\b r}\vp(t)\,,
\ee
where $\b$ is a constant and $\vp$ is a stationary solution of the diffusion equation.\footnote{The ``Wick-rotated'' diffusion equation ($r\to\rmi r$) is a Schr\"odinger equation, so $\b$ is the analogue of the energy eigenvalue of stationary solutions.}
Since $\b$ only changes the normalization of the field and its potential, we can set $\b=0$ without loss of generality:
\be\label{dit}
(\B+\a R)\vp=0\,.
\ee
Later on we will actually see that most of the solutions do require $\b=0$.

Even if the $r$ dependence is trivial, Eq.~\Eq{dit} contains a class of full-fledged dynamical solutions which are nontrivial in time and, from Eq.~\Eq{sceq}, have
\be\label{V2}
V = \frac{\vp^2}{2}\,.
\ee
The sign of the effective potential depends on the sign of $r_*\a R$:
\be\label{Weff2}
W(\vp) = \left(1-\rme^{r_*\alpha R}\right)\frac{\vp^2}{2}\,.
\ee
Unless indicated otherwise, we will choose the sign of $r_*$ so that $W$ is bounded from below.
Since the interaction is quadratic, we regard this case as perturbative in a quantum field theory sense, although it is still fully nonlocal. Notice also that, for quadratic potentials, the equation of motion is linear in $\phi$ like the diffusion equation. Only in this case does the diffusion equation contain information on the structure of the potential, which explains why one can obtain exact solutions contrary to the typical output of the general method.

The Einstein equations on stationary solutions are very simple. For any background, $\Sigma_{\mu\nu}=0$:
\ba
0&=& \left(\tfrac12+2\a\right) g_{\mu\nu} \left(\vp\B\vp+\p_\s\vp\p^\s\vp\right)-(1+2\a)\p_\mu \vp\p_\nu\vp\nonumber\\
&&+\a\vp^2 R_{\mu\nu}-2\a\vp\N_\mu\N_\nu\vp\,,\label{smn0}
\ea
where we used Eqs.~\Eq{EE1}, \Eq{V2} and $\B(\vp^2)=2\vp\B\vp+2\p_\s\vp\p^\s\vp$. After using Eq.~\Eq{dit}, the Friedmann ($00$ and trace) equations for a homogeneous field become
\ba
0&=& \left[\frac{\textsc{k}}{a^2}+\left(\frac{D}{2}-1\right)H^2\right]\vp^2-\frac{1}{2\a(D-1)}\dot\vp^2+2H\vp\dot\vp\,,\nonumber\\&&\label{fr00}\\
0&=& [4\a(D-1)+D-2]\left(\dot\vp^2+\a R\vp^2\right)\,.\label{tra}
\ea
The second equation is automatically satisfied for
\be\label{ast}
\a=-\a_*\equiv-\frac14\frac{D-2}{D-1}.
\ee
For the time being we assume $\a\neq \a_*$ and present solutions to the Einstein's equations with nontrivial $\vp(t)$ and $a(t)$ profiles.

We start with de Sitter metric,
\be\label{dS1}
a(t)=\rme^{Ht}\,,\qquad H={\rm const}.
\ee
For a flat background $\textsc{k}=0$, one finds
\be\label{dS0sol}
\vp(t) =\vp_0\rme^{-\frac12(D-1)H t}\,,\qquad \a = -\frac{D-1}{4D}\,.
\ee
The scalar field rolls down its potential towards its minimum, reaching it asymptotically. This feature is similar to what was found in \cite{cuta6} (where, however, the Einstein equations were not solved) and is typical of nonlocal models where the cosmological friction is enhanced by nonlocal operators. This is determined by the form of the nonlocal equations of motion, and it happens even in this case where the nonlocal operators are trivialized on stationary solutions.

Another class of solutions is power law:
\bs\label{pow}\ba
a(t) &=& t^p\,,\\
\vp(t) &=& \vp_0 t^q\,.
\ea\es
Inflation ($\ddot a>0$) happens when $p>1$. When $\textsc{k}=0$, the solution is
\ba
\a &=& -\frac{(Dp-p-1)^2}{4(D-1)(Dp-2)p}\,,\qquad p\neq 1\,,\nonumber\\
q &=& -\frac12(Dp-p-1)\,.\label{pls}
\ea
In four dimensions and for $p>1/3$, $q<0$ and the field rolls down towards the minimum.

We now check the classical stability of these solutions in synchronous time formalism.\footnote{For an introduction to phase space analysis, see \cite{CLW}.
Here it is worth mentioning a caveat about the choice of clocks. In general, time $t$ is an unphysical parameter and one has to choose an internal physical clock. This can be one of the matter fields in the total action, e.g., a scalar field or a barotropic fluid. Failure to do so can sometimes lead to inconsistencies in the stability analysis. The reason \cite{CL2} is that perturbations in synchronous time discriminate between trajectories differing only by a shift in time, which are actually physically equivalent and should be identified. On the other hand, in the presence of an internal clock time shifts can be physically distinguished. In the analysis below, this turns out not to be an issue. One can convince oneself by noticing that an extra dust component in the system would leave the perturbed equations unchanged.} A background solution $(H(t),\vp(t))$ is perturbed homogeneously,
\be\label{stape}
H(t)\to H(t)+\delta H(t)\,,\qquad \vp(t)\to \vp(t)+\delta\vp(t)\,,
\ee
and the equations of motion \Eq{dit}, \Eq{fr00}, and \Eq{tra} are linearized (the scalar equation of motion is an identity). The system can be written in a matrix form:
\be\label{matr}
\dot{\d X}=M\,\d X\,,
\ee
where
\be
\d X\equiv \begin{pmatrix} \d H \\ \d\vp \end{pmatrix}\,,
\ee
and the entries $m_{i\!j}=(M)_{i\!j}$ of the $2\times 2$ matrix $M$ are calculated on the background solution. The characteristic equation
\be
\det(M-\lambda \mathbbm{1})=0
\ee
determines the eigenvalues $\lambda$ and a solution is stable provided ${\rm Re}(\lambda)\leq 0$. In general, the eigenvalues are time dependent, in which case they are interpreted as evaluated at a given time $t$ \cite{CGST}. 

For all the solutions above $m_{12}=0$, so that $\lambda_1=m_{11}$ and $\lambda_2=m_{22}$. For the de Sitter solution \Eq{dS0sol},
\be
\lambda_1=0\,,\qquad \lambda_2=-\frac12(D-1)H<0\,,
\ee
thus implying stability. For the power-law solution \Eq{pls} ($t>0$),
\be
\lambda_1=-\frac{2}{t}\,,\qquad \lambda_2=-\frac{(D-1)p+1}{2t}\,,
\ee
the latter being negative when
\be
p>-\frac{1}{D-1}\,,
\ee
which is true for an expanding universe ($p>0$).

Before moving on we mention an open interpretational issue. To check the stability in a standard way, in principle one should perturb the full equations of motion and not their push-forward on stationary solutions, unless the ansatz itself [Eq.~\Eq{dit}] is stable.
Upon perturbation, the full diffusion equation stays linear in the scalar field but contains a term proportional to $\p_r\delta\phi$, which we have ignored, and a source term coming from the variation of the metric. 
The problem is that, while in local gravity the perturbed equations are linearized and so are relatively easy to solve, here they are indeed linear, but still nonlocal (nontrivial $r$ dependence).

More work might be required in the future to clarify the stability analysis of dynamical solutions. For the moment, however, notice the following. Assume that a given solution $\phi(r,t)$ of the diffusion equation, stationary or not, is a solution $\phi(t)=\phi(r_1,t)$ of the equations of motion for some $r=r_1$. In the $(r,\phi)$ plane, $\phi(t)$ is a straight line at $r=r_1$ [eventually a half-line or a segment, if the field $\phi(t)$ is bounded]. A stationary solution has no $r$ dependence, which we can fix by setting $r_1=0$. Suppose now that the linear perturbation $\delta\phi(t)=\delta\phi(r_2,t)$ solves the linearized equations for some $r_2$. This is represented by another line (or segment) parallel to the first and at distance $\Delta r=|r_2-r_1|$. For a generic \emph{nonstationary linear perturbation}, $\Delta r$ will be determined by the specific background and perturbed solutions, and is not necessarily small. On the other hand, for a \emph{stationary linear perturbation} $\Delta r=0$.

According to one interpretation, linear perturbations are nonstationary for a background solution with general self-interaction $V(\phi)$, but in the case of stationary solutions we expect that also the perturbation is stationary ($\p_r\delta\phi=0$). The main reason is that the diffusion equation is an external input which knows about the dynamics only via the background metric, but that in general is ignorant about the shape of the scalar potential. However, for stationary solutions the background potential $V$ is quadratic, which is the only case where the diffusion equation is implicitly sensitive to the scalar field dynamics. In fact, the diffusion equation can encode information about the potential by adding a mass term,
but this is a trivial normalization along the diffusion flow. Therefore, the background and perturbed diffusion equations have basically the same implicit knowledge of the scalar dynamics. The theory of diffusion suggests that the only exact solution $\delta\phi$ will be stationary as well or quasistationary ($\Delta r\sim 0$), due to the delicate dependence of the solution of the diffusion equation from the boundary conditions.

A second interpretation defines the stability as pertaining only to the spacetime dependence and not, as in nonstationary perturbations, also to the $r$ dependence. This is consistent with the idea that the $r$ dependence of the scalar is really a mathematical trick, and it is the same as imposing that the ansatz \Eq{dit} be stable by definition. In the case of nonstationary background solutions, this option imposes that the most general perturbation of the most general solution $\phi(r_1,t)$ be always of the form $\delta\phi(0,t)$.

The second interpretation prescribes that all solutions be perturbed by stationary fluctuations, while the first reaches this conclusion only for stationary solutions and only at the linear level. In the above stability analysis we have considered stationary perturbations of the form \Eq{stape} and tacitly embraced the second view.


\subsection{General solutions with quadratic potential in conformal gravity}\label{cos3}

The flat de Sitter solution can be generalized straightforwardly to any $f(R)$ theory. In fact, it is sufficient to make the replacement
\be
\a\to \a_{\rm eff}\equiv f'(R_{\rm dS})
\ee
in Eq.~\Eq{dS0sol}, where $R_{\rm dS}=D(D-1)H^2$. Depending on the form of $f$, the value of the Hubble constant is determined by $\a_{\rm eff}$.

Another generalization is based on the fact that the most general solution with $\a=-\a_*$ can be found for \emph{any} FRW background via the following shortcut. 

The critical value Eq.~\Eq{ast} is well-known in conformal gravity models \cite{FT}. In four dimensions, $\a=-1/6$. Consider a metric $g_{\mu\nu}$ and the conformal transformation
\ba
\bar g_{\mu\nu} &\equiv& \Omega^2 g_{\mu\nu}\,,\\
\bar\phi &\equiv& \Omega^{1-\frac{D}2}\phi\,,\label{fbf}
\ea
for some $r$-independent $\Omega=\Omega(x)$. The Christoffel symbols transform as
\ba
\Gamma^\lambda_{\mu\nu} &=& \bar\Gamma^\lambda_{\mu\nu}-\left[2\delta^\lambda_{(\mu}\cO_{\nu)}- \bar g_{\mu\nu}\cO^\lambda\right]\,,
\ea
where
\be
\cO_\mu\equiv \bar\N_\mu\ln\Omega,\qquad \cO\equiv \bar \N^\mu\cO_\mu=\bar\Box\ln\Omega\,.
\ee
Noting that
\ba
\Omega^{-2}R &=& \bar R+2(D-1)\cO-(D-1)(D-2)\cO_\mu \cO^\mu\,,\nonumber\\&&\\
\Omega^{-2}\B &=& \bar\B-(D-2)\cO^\mu\bar\N_\mu\,,
\ea
one can see that the combination $(\B+\a R)\phi$ has a well-defined conformal weight only if $\a=-\a_*$:
\be\nonumber
(\B-\a_* R)\phi = \Omega^2(\bar\B-\a_*\bar R)\bar\phi\,.
\ee
On a background where $\bar R=R_0={\rm const}$, the diffusion equation is equivalent to
\be\label{omdif}
\Omega^2(\bar\B-\a_*R_0+\Omega^{\frac{D}2-3}\p_r)\bar\phi=0\,.
\ee
We specialize now to FRW backgrounds, where the metric can be written in conformal time
\be
\tau\equiv \int \frac{\rmd t}{a(t)}\,,
\ee
so that $\Omega=a^{-1}$, $\bar\B=-\p_\tau^2$, and $R_0=2(D-1)\textsc{k}$. For stationary solutions, the diffusion equation reduces to Eq.~\Eq{omdif}:
\be\label{cond}
\bar\vp''+{\rm sgn}(\textsc{k})\g^2\bar\vp=0\,,
\ee
where 
\be
\g^2\equiv \left(\frac{D}{2}-1\right)|\textsc{k}|\,,
\ee
and primes denote derivatives with respect to $\tau$. Just solving this equation, one obtains a very wide class of solutions from Eq.~\Eq{fbf},
\be\label{geso}
\vp(\tau)=[a(\tau)]^{1-\frac{D}2}\bar\vp(\tau)\,.
\ee
In fact, the whole dynamics reduces to one equation, encoding both Eq.~\Eq{cond} and a constraint on $\bar\vp'$ and the curvature. To show this, instead of the trace equation \Eq{tra} one takes the $ii$-component equation $\Sigma_{ii}=0$,
\be
0= -\left(\frac12+2\a\right)\left(\dot\vp^2+\a R\vp^2\right)+\a\vp^2 \tilde R+2\a H\vp\dot\vp\,.\label{iieq}
\ee
Replacing Eq.~\Eq{geso} in Eq.~\Eq{fr00} or \Eq{iieq} with $\a=-\a_*$, one obtains just the same equation
\be\label{Kconstr}
(\bar\vp')^2+{\rm sgn}(\textsc{k})\g^2\bar\vp^2=0\,,
\ee
which can be solved only if $\textsc{k}=0$ or $\textsc{k}=-1$. There are no closed-universe real solutions. Differentiation of Eq.~\Eq{Kconstr} in conformal time yields Eq.~\Eq{cond}, so for the critical value \Eq{ast} the stationary problem is drastically simplified. This is possible only for stationary solutions, where geometry is factorized out of the equations of motion. Let us consider a few concrete examples.

\subsubsection{Flat universe ($\textsc{k}=0$)}

The $\bar g$ frame is Minkowski and $\bar\vp=1$ is a solution of the free equation $\bar\vp''=0$.\footnote{Also $\bar\vp=\tau$ solves the diffusion equation, but it does not solve Eq.~\Eq{Kconstr}.}
Then, the general solution is
\be
\vp(\tau)=[a(\tau)]^{1-\frac{D}2}\,.
\ee
For the de Sitter background \Eq{dS1} in synchronous time,
\be
\vp(t) =\rme^{-\left(\frac{D}2-1\right)H t}\,,
\ee
and the evolution of the scalar field is qualitatively the same as for Eq.~\Eq{dS0sol}.

For the power-law profile \Eq{pow}, the solution has
\be
q = -\left(\frac{D}2-1\right)p\,.\label{crit2}
\ee

Since the algorithm given by Eqs.~\Eq{cond} and \Eq{geso} is valid for any choice of the scale factor, there is an infinite set of solutions without big bang singularity. A $D=4$ bouncing solution with zero intrinsic curvature is ($r_*>0$)
\ba
&& a(t) = \frac{1}{\vp(t)} =\cosh t\,,\qquad H = \tanh t\,,\nonumber\\
&& R=6[1+(\tanh t)^2]\,.\label{cosh}
\ea
In the contracting phase the scalar field rolls from its minimum up to a maximum value. At the inversion point the universe bounces and starts expanding, while $\vp$ roll back towards the global minimum. This solution is the scalar-tensor analogue, in the sense of Eq.~\Eq{act1}, of the bouncing solution of \cite{BMS}. The fact that the scale factor $a$ is nonsingular, however, is to be ascribed more to the choice of a conformal operator rather than to the good ultraviolet properties of these nonlocal theories.

\begin{figure}\begin{center}
\includegraphics[width=8.5cm]{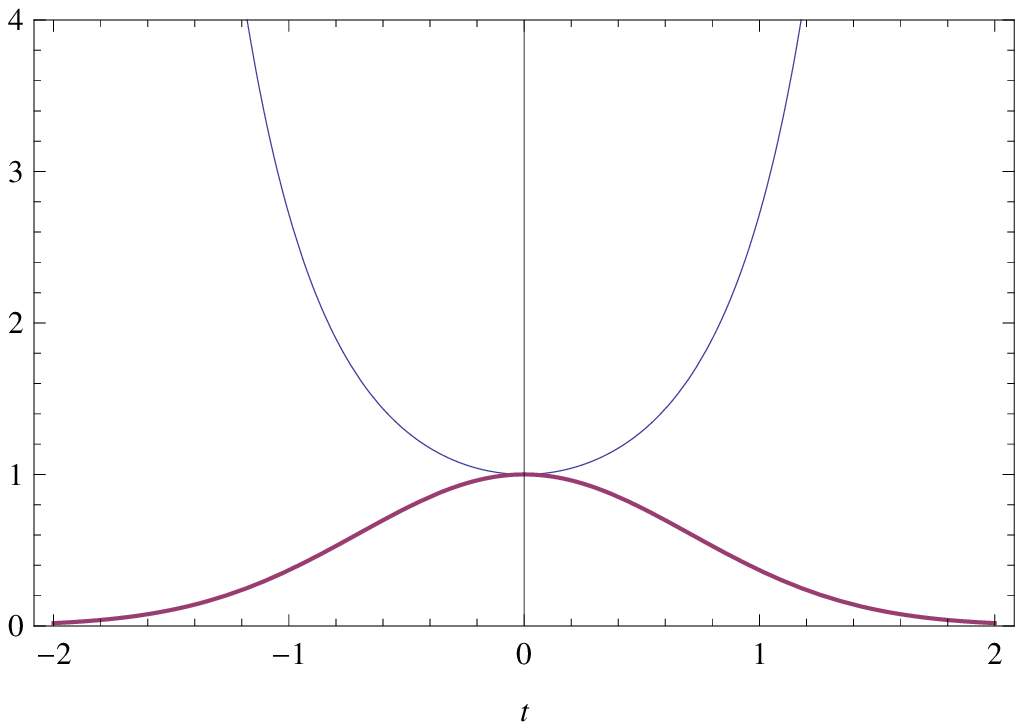}
\caption{\label{fig1} Bouncing $\textsc{k}=0$ solution \Eq{bou0} with $D=4$ and $H_0=2$. Thin line: scale factor $a(t)$. Thick line: scalar profile $\vp(t)=1/a(t)$.}
\end{center}\end{figure}

Another bouncing profile is the superaccelerating cosmology
\ba
&& a(t) = \exp\left(\frac12 H_0 t^2\right)\,,\qquad H =H_0t\,,\label{bou0}
\ea
which is plotted in Fig.~\ref{fig1} for $D=4$. The scalar field evolves from the global minimum up to the inversion point $\vp=1$ at the bounce. Then, cosmological friction drags it back to the minimum.

\subsubsection{Open universe ($\textsc{k}=-1$)}

\begin{figure}\begin{center}
\includegraphics[width=8.5cm]{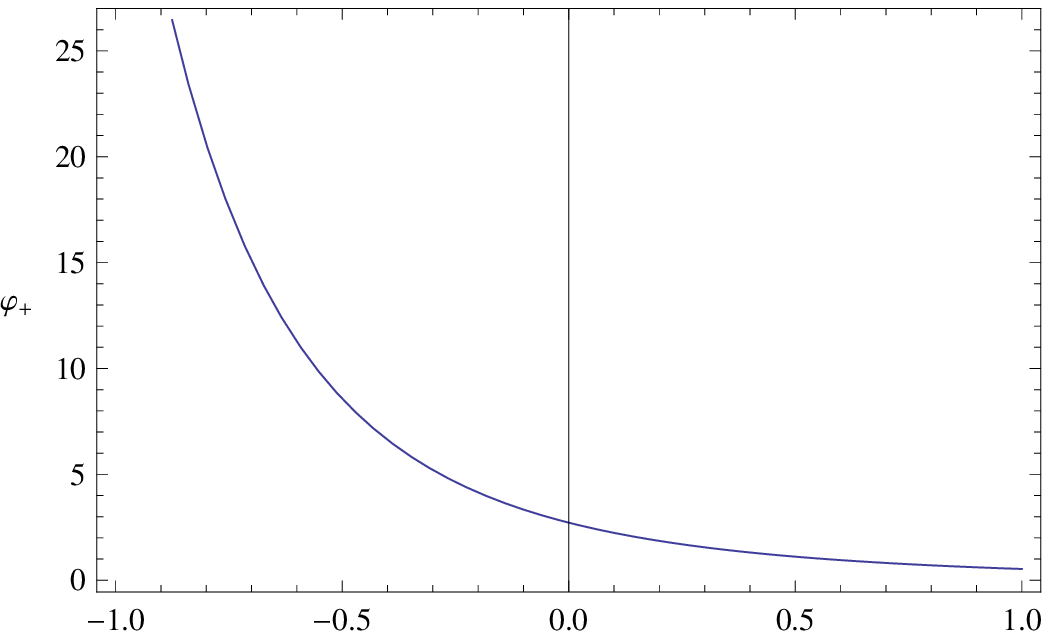}
\includegraphics[width=8.5cm]{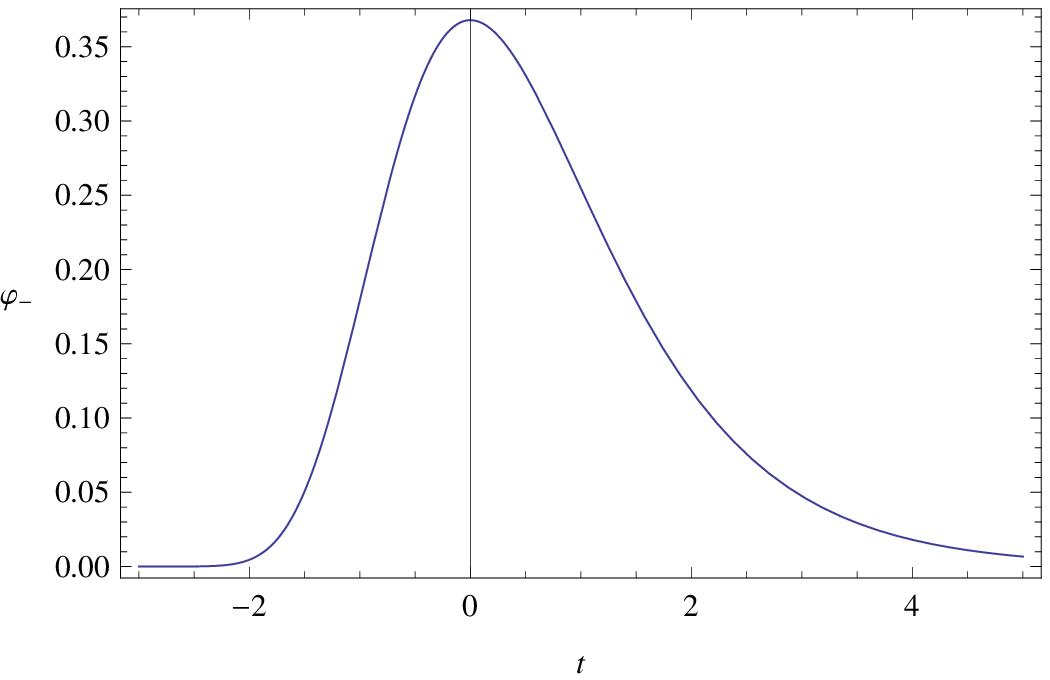}
\caption{\label{fig2}de Sitter $\textsc{k}=-1$ solutions $\vp_\pm(t)$ [Eq.~\Eq{k1ds}] with $D=4$.}
\end{center}
\end{figure}

In the open case, Eq.~\Eq{cond} is solved by exponentials and there are two general dynamical solutions which read
\be\label{vptau1}
\vp_\pm(\tau)=[a(\tau)]^{-\g^2}\rme^{\mp\g\tau}\,.
\ee
Therefore, and somehow surprisingly, there are two ``de Sitter'' ($H={\rm const}$) 
 solutions also for an open universe:
\be\label{k1ds}
\vp_\pm(t)=\rme^{-\g^2 H t}\exp\left(\pm\frac{\g}{H}\,\rme^{-Ht}\right)\,,
\ee
where we used $\tau=-\rme^{-Ht}/H$. The profile $\vp_+$ has the same features as the flat solutions and rolls towards the global minimum. The solution $\vp_-$ climbs from the minimum up to some maximum value, then rolls back down (Fig.~\ref{fig2}).

\begin{figure}\begin{center}
\includegraphics[width=8.5cm]{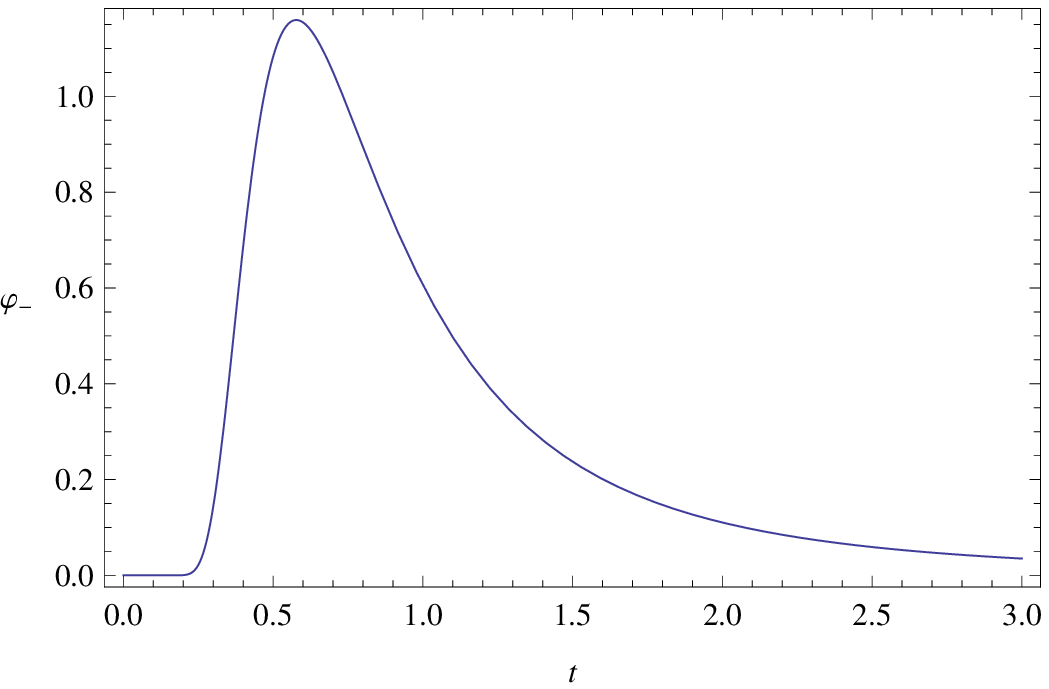}
\caption{\label{fig3} Power-law $\textsc{k}=-1$ solution $\vp_-(t)$ [Eq.~\Eq{powk1}] with $D=4$ and $p=3$.}
\end{center}
\end{figure}

For the power-law profiles \Eq{pow}, when $p=1$ (linear scale factor) the solutions have
\be\label{k1pl}
q_\pm=-\sqrt{\frac{D}2-1}\left(\sqrt{\frac{D}2-1}\pm1\right)\,.
\ee
In four dimensions, only the positive root is nontrivial, $q=-2$. For arbitrary $p$, the solutions are
\be\label{powk1}
\vp_\pm(t)=t^{-\g^2 p}\rme^{\pm\frac{\g}{p-1} t^{1-p}}\,.
\ee
The typical plot of, say, $\vp_-$ is shown in Fig.~\ref{fig3} and is similar to the previous one. For comparison, the exact de Sitter and power-law solutions in standard general relativity are reported in the Appendix.

\begin{figure}\begin{center}
\includegraphics[width=8.5cm]{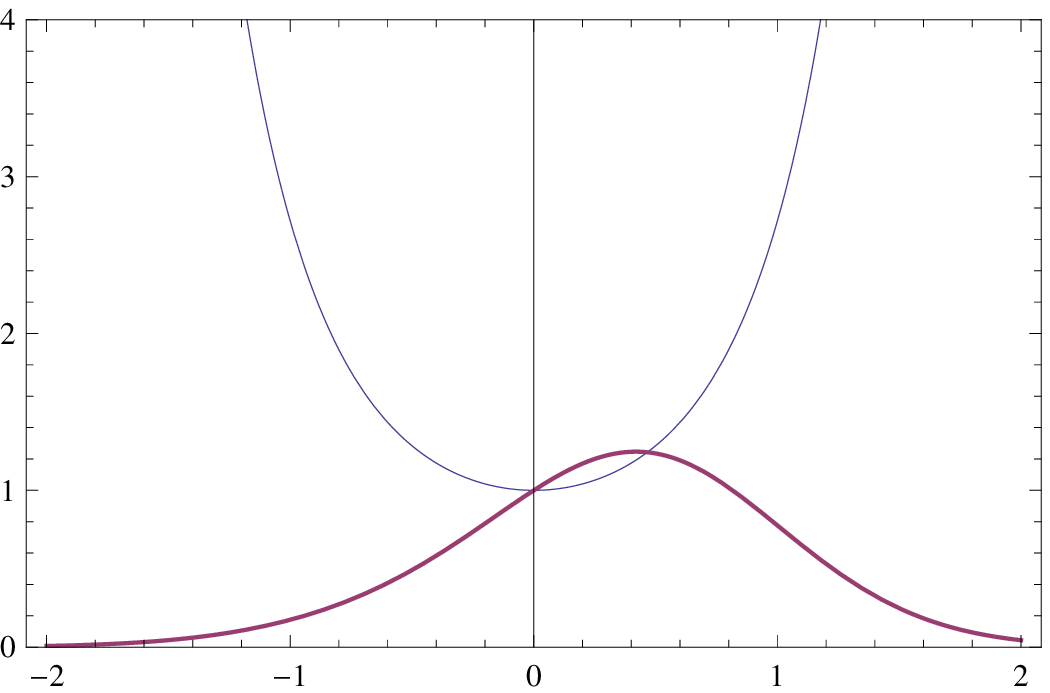}
\caption{\label{fig4} Asymmetric bouncing solution given by Eqs.~\Eq{vptau1} and \Eq{bou1}, with $D=4$ and $H_0=2$. Thin line: scale factor $a(t)$. Thick line: scalar profile $\vp_-(t)$.}
\end{center}
\end{figure}

For the bouncing solution Eq.~\Eq{bou0}, the conformal time is
\be\label{bou1}
\tau=\sqrt{\frac{\pi}{2H_0}} {\rm erf}\left(\sqrt{\frac{H_0}{2}}t\right)\,,
\ee
thus leading to the asymmetric lump $\vp_-(t)$ shown in Fig.~\ref{fig4}. The evolution is similar to the flat case, except that now open geometry helps the scalar field to climb up to a higher inversion point before being dragged back by cosmic superfriction. The lump $\vp_+(t)$ has its maximum at $t<0$. This corresponds to a solution with a different initial condition for the scalar field: $\vp_+$ rapidly reaches the (higher) inversion point before the bounce.

\begin{figure}\begin{center}
\includegraphics[width=8.5cm]{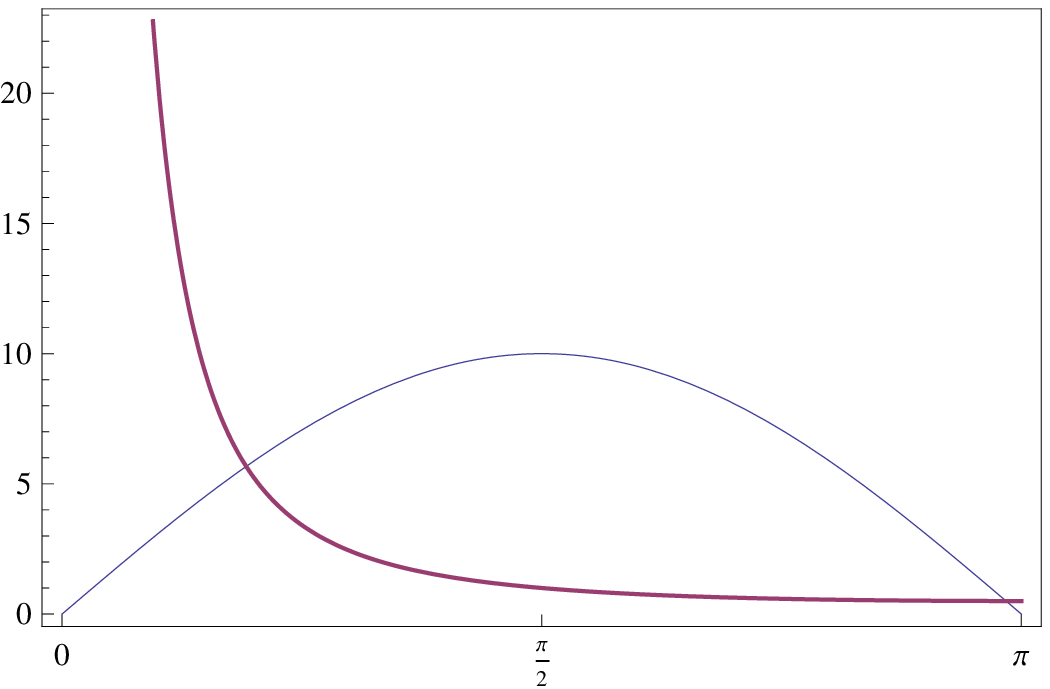}
\includegraphics[width=8.5cm]{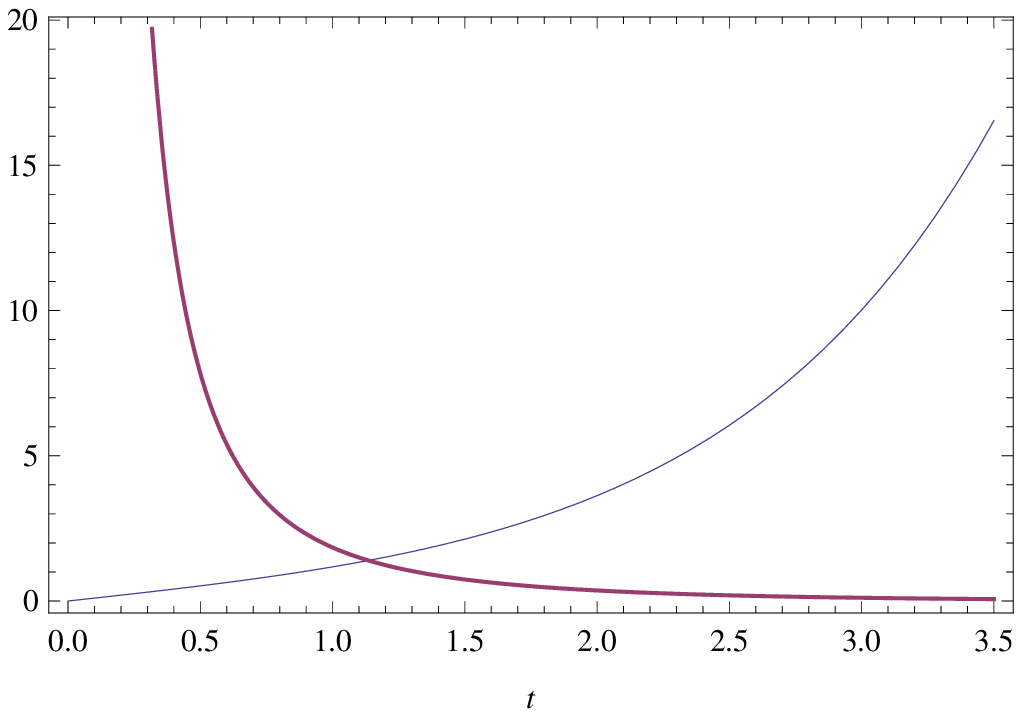}
\caption{\label{fig5} Upper panel: scale factor $10 a(t)$ (thin line) and scalar profile $\vp(t)$ (thick line) for the recollapsing solution Eq.~\Eq{sin}. Bottom panel: scale factor $a(t)$ (thin line) and scalar profile $\vp(t)$ (thick line) for the expanding solution Eq.~\Eq{sinh}.}
\end{center}
\end{figure}

Still in an open universe, we sketch some other solutions in four dimensions which well illustrate the exotic properties which can emerge in a nonlocal conformal setting. The first case is periodic in time, so we restrict it to half a period with positive scale factor  ($r_*<0$):
\ba
a(t) &=& \sin t\,,\qquad t\in [0,\pi]\,,\nonumber\\
\textsc{k} &=& -1\,,\qquad R=-12\,,\label{sin}\\
\vp(t) &=& \frac{1}{1-\cos t}\,,\nonumber
\ea
which is shown in the upper panel of Fig.~\ref{fig5}. Despite being open, the universe recollapses onto itself while the scalar field rolls towards the global minimum $\vp=0$. Eventually one hits a future big crunch singularity at finite times. Another solution consists in a ``Wick rotation'' of the former ($r_*>0$): 
\ba
&& a(t) = \sinh t\,,\nonumber\\
&& \textsc{k} = -1\,,\qquad R=12\,,\label{sinh}\\
&& \vp(t) = \frac{1}{\cosh t-1}\,,\nonumber
\ea
depicted in the bottom panel of Fig.~\ref{fig5}.

The usual stability analysis does not apply to the above solutions because $a(t)$ is factorized out of the equations of motion. Thus, all points in the phase space plane $(\dot a,\dot\vp)$ are fixed points and there are no attractors. This is a rather bizarre situation in cosmology. Typically, a solution of the background dynamics is probed by perturbing it slightly in all sectors, inclusive gravity, as done previously. Thus one checks whether one will hit the same solution after evolving the system from slightly different initial conditions. In the conformal system with $\a=-\a_*$, on the other hand, this operation is ill defined because $a(t)$ is a fixed input. 

Can we conclude that the scale factor is nondynamical and, thus, the conformal solutions of this section are not physically sensible? We argue in favor of a negative answer. Nonlocal models are notoriously rigid and the choice of initial conditions is much more restricted with respect to local theories with a standard Cauchy problem \cite{cutac,cuta2,MZ}. Here we see this property in action, with a further restriction of the background choice. Moreover, the scale factor factorizes out only because we have assumed trivial diffusion, Eq.~\Eq{cond}. A realistic stability analysis should go beyond the homogeneous level, where the solutions are no longer stationary in the diffusion sense; this would correspond to looking at semiclassical (inhomogeneous) perturbations.


\subsection{Asymptotic solutions and spontaneous symmetry breaking}\label{cos4}

We have found exact, cosmologically nontrivial solutions which are constant along the diffusion direction but one might ask whether there exist also nonstationary diffusing profiles. This is a rather difficult question to answer because there is no systematic method to solve the equations of motion starting from a solution of the diffusion equation. In this paper we will limit ourselves to some general remarks for $f(R)=\a R$.

The diffusion equation transfers the nonlocal degrees of freedom into the extra direction \cite{cuta3,cuta7} and one can study the system locally in spacetime coordinates without risk of falling into the contradictory situations typical of nonlocal dynamics. In particular, it is possible to expand field profiles and equations in series and look for asymptotic solutions.

The static potential felt by the scalar particle is, according to Eq.~\Eq{act1},
\be\label{Weff}
W(\phi) = V(\phi)- \frac12\rme^{r_*\alpha R} \phi^2\,.
\ee
The curvature term matches a possible quadratic term $m^2\phi^2/2$ in $V$ and the net result is a dynamical rescaling of the squared mass, which may be even negative (tachyon). A particularly important situation is when $V$ has a minimum at the origin with a power higher than 2 (e.g., $V\sim\phi^4$ but it could be of any other form). In this case, the dynamical term of the potential always converts the minimum $\phi=0$ of $V$ into a local maximum of $W$, and the minima are located at $\phi_0\neq 0$. Classically, stable solutions should tend to these minima,
\be\label{asico}
\phi\ \stackrel{t\to\infty}{\longrightarrow} \ \phi_0\,.
\ee
Quantum mechanically, the scalar field is expected to undergo spontaneous symmetry breaking and take a nonvanishing expectation value on the vacuum, $\langle \phi \rangle_0 = \phi_0$. In this case the system relaxes to pure gravity, Eq.~\Eq{gravo}, the expectation value of the scalar field playing the role of Newton's constant. This is precisely the same kind of mechanism that happens in string theory with the dilaton field.

If the scalar field potential is to be determined \emph{a posteriori}, in the presence of gravity a convenient approach is to factorize $V$ out and consider the equation of motion
\be\label{conbi}
0=\int_0^{r_*}\rmd s (D\Sigma_{00}+\Sigma)\,,
\ee
which can be solved for given FRW profiles. In turn, this determines the scalar field profile in the diffusion equation.
Looking for asymptotically (nonvanishing) constant solutions for the matter field in 
 Eq.~\Eq{conbi}, one gets $DR_{00}+R=0$, that is,
\be
\label{geome}
a^2\dot H = \frac{2\textsc{k}}{D-2}\,.
\ee
This constrains the asymptotic geometry of spacetime. When $\textsc{k}=0$, the only solution is de Sitter. For $\textsc{k}=\pm1$ and $D=4$ the scale factor is, respectively, $a=(\cosh b t)/b$ and $a=(\sinh b t)/b$, where $b$ is an arbitrary constant.\footnote{Clearly, the symmetry breaking argument does not apply to the corresponding exact stationary solutions with quadratic potential, Eqs.~\Eq{dS0sol}, \Eq{cosh} and \Eq{sinh}.} The curvature is always a positive constant, $R_0=12 b^2$. In the limit $b\to 0$, the $\textsc{k}=-1$ solution reduces to the linear case $a=t$. This is the only  nontrivial solution of Eq.~\Eq{geome} with asymptotically vanishing curvature $R_0$. Whatever the full solutions are, they can only be constructed by considering the above scale factors as either exact or asymptotic profiles.

We conclude this part by determining the relations between the constants defining the spacetime geometry ($\Lambda$ and $G$) and those coming from matter [$V_0=V(\phi_0)$ and $\phi_0$].
This can be achieved in two independent ways, by looking either at the action Eq.~\Eq{act1} or at the equations of motion, Eqs.~\Eq{EE1} and \Eq{EE2}. Taking the weak nonlocal limit of the action ($r_*\ll 1$), we get the standard Einstein--Hilbert action with cosmological and Newton constants given by 
\be\label{constants}
\Lambda_\phi= \frac{1}{2 r_* \alpha} \left(\frac{2V_0}{\phi_0^2}-1\right)\,, \qquad   8 \pi G_\phi=\frac{1}{r_* \alpha\phi_0^2}\,.
\ee 
By consistency, we expect Eqs.~\Eq{EE1} and \Eq{EE2} to reproduce the Einstein equations in vacuum with the values \Eq{constants}. To this aim, it is important to note that the asymptotic value of the scalar field depends on $r_*$. The potential $W$ is explicitly $r_*$ dependent and, as a consequence, so are its local minima.

Once again, we can appeal to the diffusion equation governing the flow in the $r_*$ variable. In fact,
\ba
\phi(r-r_*,x) &=& \phi(r,x)-r_* \partial_r \phi(r,x)+O(r_*^2)\nonumber\\
&=&[1+r_*\B+r_*\a R+O(r_*^2)]\phi(r,x).\label{phir0}
\ea
One could have just ignored the diffusion picture and considered a small $r_*$ expansion in $\phi(1-r_*,x)=\rme^{r_*(\B+\a R)}\phi(1,x)$,
\be
\phi(1-r_*,x) = [1+r_*\B+r_*\a R+O(r_*^2,\B^2)]\phi(1,x)\,.
\ee
However, in the latter expression it is not clear whether we are entitled to safely take an asymptotic limit $\phi_{\rm asymp}$ for $\phi$, as we have to know the contribution of \emph{all} the derivatives of the field. On the other hand, the diffusion method regards $\phi(r,x)$ as a local field with two arguments, and the small $r_*$ expansion is a genuine local expansion. In other words, the diffusion picture justifies manipulations of asymptotic solutions in the nonlocal model.

Taking the asymptotic limit $x\to \infty$ of Eq.~\Eq{phir0}, we get
\be\label{phir}
\lim_{x\to \infty}\phi(1-r_*,x)= [1+r_*\a R_0 + O(r_*^2)]\phi_0\,.
\ee
Performing the asymptotic limit in the equations of motion \Eq{EE1} and \Eq{EE2}, expanding in $r_*$ up to the first order, and taking Eq.~\Eq{phir} into account, one easily gets the standard Einstein equations in vacuum, with Newton and cosmological constants as in Eq.~\Eq{constants}. 

Another interesting avenue to explore for nonstationary cosmological solutions is the conformal case $\a=-\a_*$. Then, the diffusion equation becomes
\be\label{cond2}
\bar\phi''+{\rm sgn}(\textsc{k})\g^2\bar\phi-a^{3-\frac{D}2}\p_r\bar\phi=0\,.
\ee
Interestingly, in six dimensions this is the usual homogeneous diffusion equation in conformal time, of which we know the solutions. For example, on a de Sitter background the Gaussian profile $\bar\phi(r,\tau)$ becomes an asymmetric kink in synchronous time, while for the power-law expansion it is a very flat lump. One should check, however, if the profiles $\phi$ solve, even asymptotically, the Einstein equations for a given potential. This is not guaranteed. In fact, unlike the case of stationary solutions the system is not trivialized to Minkowski. The reason is simple. Let
\be
\phi_n\equiv (\B-\a_* R)^n\phi\,.
\ee
The field $\phi_n$ is conformal with weight $a^{-2n}$ when $n=(D/2-1)/2$. Hence, none of the $\phi_n$ is conformal (except possibly one if $D=6+2k$, $k\in\mathbb{N}$) and the object
\be
\rme^{\B-\a_* R}\phi = \sum_{n=0}^\infty \frac{\phi_n}{n!}
\ee
contains an infinite number of $a$ factors, which survive in $\Sigma_{\mu\nu}$.


\subsection{Inclusion of cosmological fluids}\label{matte}

So far we have been focusing on solutions where the only matter component was the nonminimally coupled scalar field. The first reason to do so was simplicity; the structure of the above solutions is already rich enough in the stationary ``empty'' (no fluid) regime to illustrate some features of the class of nonlocal theories under inspection. Second, as mentioned in the introduction, we expect nonlocality to manifest itself in the early universe, in particular, through a bounce or during inflation when dust and radiation are typically (but not always) ``switched off.'' Third, our purpose was to define a nonlocal theory which would be consistently nonlocal in all sectors. This means that it is not clearly consistent to introduce a traditional cosmological fluid, because its action is intrinsically local. Instead, nonlocality can be easily and rigorously implemented in a field theory and, if one were interested in building a realistic multicomponent postinflationary cosmology, one should define a set of fields with asymptotically perfect-fluid-like equations of state.

Of course one can just ignore this issue, add ordinary fluids, and study their cosmology numerically, eventually placing observational constraints on the parameters of the theory. This exercise lies beyond the above mentioned foundational framework, but in this section we can make some general comments on analytic single-fluid solutions.

Let us add to the action \Eq{act1} a generic matter part $S_{\rm m}$ giving rise to the energy-momentum tensor $T_{\mu\nu}=-(2/\sqrt{-g})\delta S_{\rm m}/\delta g^{\mu\nu}$. The latter constitutes the left-hand side of the Einstein equations \Eq{EE1}. For an ordinary perfect fluid with energy density $\rho$ and pressure $p$, the energy-momentum tensor is $T_{\mu\nu}=(\rho+p)\,u_\mu u_\nu+p\,g_{\mu\nu}$, where $u^\mu$ is the comoving four-velocity field (unit timelike vector, $u_\mu u^\mu=-1$) tangent to a fluid element's worldline. The continuity equation reads $\dot\rho+(D-1)H(\rho+p)=0$, while Eqs.~\Eq{fr0} and \Eq{tr0} become
\ba
&&\rho = \frac12 \phi(1)\phi(1-r_*)-V[\phi(1)]+\int_0^{r_*} \rmd s\,\Sigma_{00}(s)\,,\label{fr0m}\\
&&(D-1)p-\rho =-D\left\{\frac12 \phi(1)\phi(1-r_*)-V[\phi(1)]\right\}\nonumber\\
&&\qquad\qquad\qquad\quad+\int_0^{r_*} \rmd s\,\Sigma(s)\,.\label{tr0m}
\ea
On stationary solutions, the Einstein equations are
\ba
T_{\mu\nu}&=& \left(\frac12+2\a\right) g_{\mu\nu} \left(\vp\B\vp+\p_\s\vp\p^\s\vp\right)\nonumber\\
&&-(1+2\a)\p_\mu \vp\p_\nu\vp+\a\vp^2 R_{\mu\nu}\nonumber\\
&&-2\a\vp\N_\mu\N_\nu\vp\,,\label{smn02}
\ea
where we set $r_*=1$ in the trivial integration in the right-hand side. In particular, the Friedmann equations are
\ba
&&\rho =\a(D-1) \left[\frac{\textsc{k}}{a^2}+\left(\frac{D}{2}-1\right)H^2\right]\vp^2-\frac{1}{2}\dot\vp^2\nonumber\\
&&\qquad+2\a(D-1)H\vp\dot\vp\,,\label{fr00m}\\
&&\rho-(D-1)p = \left[2\a(D-1)+\frac{D}{2}-1\right]\left(\dot\vp^2+\a R\vp^2\right)\,.\nonumber\\\label{tra2}
\ea
For a barotropic fluid, the equation of state is $p=w\rho$, where $w$ is constant. Then, from the continuity equation it follows that $\rho=\rho_0 a^{-(D-1)(w+1)}$. For pressureless dust $w=0$, while for radiation $w=1/(D-1)$. We would like to see how some of the previously found solutions are modified in this setting. We take $\rho_0>0$ and $-1<w<1$.

As before, let us first assume $\a\neq -\a_*$. For the de Sitter metric \Eq{dS1} and a flat background, the solution \Eq{dS0sol} is replaced by
\ba
\vp(t) &=&\vp_0\rme^{-\frac12(D-1)(w+1)H t}\,,\nonumber\\
 \a &=& -\frac{(D-1)(1-w^2)}{4D}\,,\label{dS0sol2}
\ea
where
\ba
&&\vp_0=\pm\frac{2}{(D-1)H}\sqrt{\frac{D\rho_0}{-w(w+1)[1+(D-1)w]}}\,,\nonumber\\
&&-\frac{1}{D-1}<w<0\,.
\ea
Neither radiation nor dust are solutions. The scalar field rolls down the minimum of the potential but at a slower rate with respect to the no-fluid case, by a factor of $1+w$.

The $\textsc{k}=0$ power-law solution \Eq{pow} and \Eq{pls} now becomes
\ba
\a &=& -\frac{(1-w)[(D-1)(w+1)p-2]}{4(Dp-2)}\,,\nonumber\\
q &=& 1-\frac12(D-1)(w+1)p\,,\label{pls2}
\ea
with
\begin{widetext}
\be
\vp_0=\pm2\sqrt{\frac{(Dp-2)\rho_0}{[1-(D-1)wp][(D-1)wp+p-2][(D-1)(w+1)p-2]}}\,.
\ee
\end{widetext}
The solution is real only for a certain range of the parameters. For instance, there are no real solutions for radiation, $w=1/(D-1)$, while for $w=0$ it must be either $p>2$ or $2/D<p<2/(D-1)$. Notice that now, in four dimensions, $q<0$ if $p>2/[3(w+1)]$. In order for the scalar field to roll towards its minimum, and assuming $0\leq w<1$, $p$ must be larger than the value $1/3$ obtained in the case without fluid.

The conformal case $\a=-\a_*$ is straightforward. The trace equation vanishes identically only in the presence of a radiation component, $w=1/(D-1)$. Then, both the $00$ and $ii$ components of the Einstein equations reduce to
\be\label{Kconstr2}
(\bar\vp')^2+{\rm sgn}(\textsc{k})\g^2\bar\vp^2+2\rho_0=0\,,
\ee
which generalizes Eq.~\Eq{Kconstr}. The flat background no longer admits solutions and only open-universe profiles survive. Contrary to Eq.~\Eq{vptau1}, the general solution is a nontrivial superposition of the solutions of the diffusion equation:
\be\label{vptau12}
\vp(\tau)=\vp_0[a(\tau)]^{-\g^2}\left(\rme^{\g\tau}+\frac{\rho_0}{2\g^2\vp_0^2}\,\rme^{-\g\tau}\right)\,,\qquad \textsc{k}=-1\,.
\ee
As before, the ``de Sitter'' ($H={\rm const}$) and power-law solutions are qualitatively similar. The scalar always rolls towards the global minimum, monotonically if $\rho_0/\vp_0^2\ll 1$ and with a smooth, temporary turnaround if $\rho_0/\vp_0^2\gg 1$.


\section{Stringlike action and cosmology}\label{slac}

The model studied so far is based on a gravity extension of a $p$-adic-like scalar field action, which has a trivial local limit $r_*\to 0$ (no dynamical degrees of freedom). We can now make a slight but crucial modification of the kinetic operator $\rme^\cK$ into a transcendental expression of the form $\rme^\cK\cK$:
\be\label{actt}
S=\int d^Dx\sqrt{-g}\left[\frac12 \phi \rme^{r_*[\B+f(R)]}\tphi-V(\phi)\right]\,,
\ee
where
\be
\tphi\equiv [\B+f(R)]\phi\,.
\ee
The linear case $f(R)=\a R$ is important for several reasons already illustrated in \cite{BMS}. First, when $r_*<0$ one formally recovers the effective spacetime action of the tachyon and the graviton in closed SFT. When gravity is switched off, Eq.~\Eq{actt} reads
\be\label{acttac}
S\sim\int d^Dx\left[\frac12 \phi \rme^{-|r_*|\B}\B\phi-V(\phi)\right]\,,
\ee
where $\phi$ is (up to a $\phi\tphi$ negative mass term) the tachyon field ``dressed'' with an exponential operator (see, e.g., \cite{cuta5} for details) and $V$ is a polynomial potential. On the other hand, when $\phi=1$ we get
\be\label{actgrav}
S\sim\int d^Dx\sqrt{-g}\,\left[\rme^{-|r_*|(\B+\a R)}R-2\Lambda\right]\,.
\ee
This is an effective nonlocal gravitational action in terms of Riemann invariants. In closed SFT we do not have any such thing because the effective level-truncated action for $g_{\mu\nu}$ is obtained in a nondiffeomorphism-invariant form \cite{KS3,Moe3}.\footnote{Nonetheless, diffeomorphisms are still part of the symmetry group of the theory \cite{KS3}.} However, by expanding Eq.~\Eq{actgrav} at linear order around Minkowski spacetime, one indeed gets the correct  propagator $\tilde G(k^2)$ in momentum space for the graviton, as shown in \cite{BMS}:
\be
\tilde G(k^2)\sim -\frac{\rme^{-|r_*|k^2}}{k^2}\,.
\ee
This propagator is ghost free and realizes asymptotic safety at large momenta, two properties expected in a ultraviolet-finite nonperturbative theory of quantum gravity \cite{cutac,cuta4,BMS}.\footnote{This type of propagator appears also in superrenormalizable quantum field theories \cite{efi77} and in the context of noncommutative theories where a minimal length is effectively induced \cite{SSN,NiR}.} Therefore, Eq.~\Eq{actt} is interesting both as a toy model for the spacetime effective dynamics of gravity in closed SFT and as a nonperturbative ansatz for the gravitational action at high energies/large curvature, which can play a major role during the very early universe.

From Eq.~\Eq{actt}, the scalar equation of motion $\delta S/\delta\phi=0$ is
\be\label{sceqt}
\tphi(1-r_*,x)=V'[\phi(1,x)]\,,
\ee
while the Einstein equations read
\ba
0 
 &=& -g_{\mu\nu}\left\{\frac12 \phi(1)\tphi(1-r_*)-V[\phi(1)]\right\}\nonumber\\
 &&+\Sigma_{\mu\nu}(0)+\int_0^{r_*} \rmd s\,\tilde\Sigma_{\mu\nu}(s)\,,
\ea
where $\tilde\Sigma_{\mu\nu}$ is $\Sigma_{\mu\nu}$ with $\phi(1-r_*+s)$ replaced by $\tphi(1-r_*+s)$ (or the symmetrized expression). With obvious notation, the trace equation is
\ba
0&=&-D\left\{\frac12 \phi(1)\tphi(1-r_*)-V[\phi(1)]\right\}+\Sigma(0)\nonumber\\
 &&+\int_0^{r_*} \rmd s\,\tilde\Sigma(s)\,.
\ea
The equations of motion could have been obtained also by replacing $V\to r_* V$ in the equations of the $p$-adic-like case, Eqs.~\Eq{sceq} and \Eq{EE1}, and differentiating with respect to $r_*$. One notices, in fact, that $\p_{r_*}\phi(1-r_*+s)=\tphi(1-r_*+s)$ and that the contribution of the upper extremum of the integral, $\Sigma_{\mu\nu}(r_*)$, is equal to $\Sigma_{\mu\nu}(0)$ upon symmetrization (i.e., splitting $\int \Sigma$ in two and replacing $s\to r_*-s$ in one of the pieces). 

To get stationary solutions with $f(R)=\a R$, it is necessary to reinstate the normalization constant $\b$ in the diffusion equation,
\be\label{ditb}
\tilde\vp+\b\vp=0\,,
\ee
which is fixed by Eq.~\Eq{sceqt}:
\be\label{beta}
V=\frac{m^2}2\, \rme^{2\b}\vp^2\,,\qquad \b= -m^2\rme^{\b r_*}\,.
\ee
The equations of motion are still $\Sigma_{\mu\nu}=0$ but with $0\neq\b\neq 1$. The Friedmann equations read
\ba
0&=& \left[\frac{\textsc{k}}{a^2}+\left(\frac{D}{2}-1\right)H^2+\frac{\b}{2\a(D-1)}\right]\vp^2\nonumber\\
 &&-\frac{1}{2\a(D-1)}\dot\vp^2+2H\vp\dot\vp\,,\\
0&=& [4\a(D-1)+D-2]\left(\dot\vp^2+\a R\vp^2\right)\nonumber\\
 &&+[4\a(D-1)+D]\b\vp^2\,.
\ea
The requirement $\b\neq 0$ is very stringent, $\a\neq-\a_*$, and the only exact solution in common with the ``$p$-adic'' case is the flat ($\textsc{k}=0$) de Sitter profile \Eq{dS1} and \Eq{dS0sol} with
\ba
&&\b     = -\frac{\a(4D\a+D-1)[4\a(D-1)+D-2]}{(4\a+1)^2}H^2\,,\nonumber\\
 &&\\
&&\vp(t) = \exp\left(\frac{2\a H}{4\a+1} t\right)\,.
\ea
The solution is stable only for certain values of $\a$. In fact, the eigenvalues of the characteristic equation for the linearized system \Eq{matr} are
\be
\lambda_1=\left(\frac{1}{4\a+1}-D\right)H\,,\qquad \lambda_2=\frac{2\a H}{4\a+1}\,.
\ee
The solution is stable for
\be
-\frac{D-1}{4D}<\a<0.
\ee
In the presence of a barotropic fluid, there is a richer structure of exact solutions. The equations of motion read
\ba
&&\rho = \a(D-1)\left[\frac{\textsc{k}}{a^2}+\left(\frac{D}{2}-1\right)H^2+\frac{\b}{2\a(D-1)}\right]\vp^2\nonumber\\
 &&\qquad-\frac{1}{2}\dot\vp^2+2H\a(D-1)\vp\dot\vp\,,\\
&&\hspace{-3mm}2[1-(D-1)w]\rho = [4\a(D-1)+D-2]\left(\dot\vp^2+\a R\vp^2\right)\nonumber\\
 &&\qquad\qquad\qquad\qquad+[4\a(D-1)+D]\b\vp^2\,.
\ea
The scalar field profile of the de Sitter solution now is
\be
\vp(t) =\vp_0\rme^{-\frac12(D-1)(w+1)H t}\,,
\ee
with
\ba
\a &=& -\frac{(D-1)^2(w+1)+\frac{4\rho_0}{H^2\vp_0^2}}{4(D-1)[D+(D-1)w]}\,,\\
 \b &=&-\frac14(D-1)[4D\a+(D-1)(1-w^2)]H^2\,.\nonumber
\ea
There are no constraints on $w$ since $\a$ diverges at the exotic value $w=-D/(D-1)<-1$.

The extra fluid terms compensate those proportional to $\b$ and allow us to consider also the conformal case $\a=-\a_*$. The equations of motion are solved for
\ba
\vp_\pm(t) &=& \vp_0[a_\pm(t)]^{-\frac12(D-1)(w+1)}\,,\nonumber\\ 
\b &=& \frac{[1-(D-1)w]\rho_0}{\vp_0^2}\,,\label{uff1}
\ea
where the scale factor is
\ba
a_\pm(t) &=& \frac{\rme^{\pm At}+4(D-2)\textsc{k}\vp_0^2\rme^{\mp At}}{4\sqrt{2[-1-(D-1)w]\rho_0}}\,,\nonumber\\ 
A &=& \sqrt{\frac{\rho_0}{-1-(D-1)w}}\,\frac{2}{\vp_0}\,.\label{uff2}
\ea
Notice that $w<-1/(D-1)$ and neither dust nor radiation are real solutions. When $\textsc{k}=0$, the solution $a_\pm$ reduces to de Sitter. Setting $\vp_0^2=1/[4(D-2)]$, the other two cases are easily written. The open-universe solution is $a\propto \sinh At$ and expands monotonically, while for $\textsc{k}=1$ there is a bounce, $a\propto\cosh At$. For other values of $\vp_0$ the bounce is asymmetric.


\section{Discussion}\label{disc}

In this paper we have constructed and solved, on cosmological backgrounds, an effective nonlocal model of gravity nonminimally coupled with a scalar field. The actions \Eq{act1} and \Eq{actt} are nonperturbative both in the order of curvature invariants,\footnote{Actions of the form \Eq{loca} are sometimes dubbed ``exponential gravity'' \cite{CENOS,Lin09}.}
\be\label{loca}
\rme^{R}\sim 1+R+\frac12R^2+\dots\,,
\ee
and in the number of derivatives acting on the metric,
\be\label{trunc}
\rme^\B R\sim R+\B R+\frac12\B^2R+\dots\,.
\ee 
Nonlocal cosmology is radically different from higher-order cosmological models of $f(R)$, Gauss--Bonnet, or $f(\textrm{Gauss--Bonnet})$ gravity. This is because truncation of a nonlocal model in spacetime derivatives produces an order $n$ Ostrogradski-like problem with altogether different physical properties. The well-known fact that theories with an infinite number of derivatives are not the large $n$ limit of finite-order actions has been also invoked to question the relevance of higher-order cosmologies in the early universe \cite{CDD}. Rather than a finite-order truncation of the gravitational action, near the big bang curvature effects should be consistently taken into account only within a fully nonperturbative framework in the sense of the left-hand side of Eqs.~\Eq{loca} and \Eq{trunc}. The right-hand side of Eq.~\Eq{trunc} might not be even well defined on a general nonlocal solution of the system \cite{cuta2}. The diffusion equation method allows one to deal with the full nonlocal operators and bypass the problems of a series expansion.

The diffusion structure we have explored is asymmetric in the gravity and scalar sector; in fact the former does not diffuse at all. The only solutions we have been able to find do not diffuse even in the matter sector (more precisely, they are stationary along the diffusion flow), but in general a nonstationary diffusion structure is necessary to solve the system with a self-interacting (higher-order potential) scalar field. For the purpose of finding analytic solutions, this should exclude the \emph{a priori} assumption that, preferring a ``symmetric'' formulation of the model, also the scalar sector does not diffuse. In this case, geometry through the curvature term $f(R)$ would replace diffusion along $r$. Therefore, the theory of diffusion associated with nonlocal actions would be simply defined differently: Diffusion always takes place through geometry, but in the case of trivial geometry (Minkowski background), this is realized by an auxiliary higher-dimensional structure. If this was really the case, however, it would be probably difficult to find analytic or semianalytic solutions with nonlinear self-interaction (nonquadratic $V$).

The actions we have studied are structurally similar to the one advanced in \cite{BMS} for the following reason. On one hand, the proposal of \cite{BMS} aimed at an ultraviolet-finite action for quantum gravity which would address the big bang singularity problem. On the other hand, we wanted an action which would be nonlocal in both matter and gravity sectors and be endowed with a diffusion structure allowing one to reduce the dynamics to a set of local equations with both a second-order differential structure (in spacetime) and an algebraic structure (in the diffusion direction \cite{cuta2,cuta3,cuta5}). These questions, however, are implicitly related: ghost and asymptotic freedom are determined by the specific choice of pseudodifferential operators, in this case one with a natural diffusion structure. So diffusion and good ultraviolet properties are tied together, as expected in string field theory \cite{cuta7}.

There is, anyway, a caveat in this comparison. We not only stressed the importance of solving a fully nonlocal action with both gravity and matter cosmological nontrivial profiles, but in doing so it was also shown how these profiles can differ, even considerably, with respect to local scenarios.\footnote{Deviations from local cosmology is not limited to background solutions. It would be interesting to study the inflationary spectra stemming from the inhomogeneous perturbation of the Einstein equations.} The simplest cosmological profiles (de Sitter and power law) are exact solutions of the nonlocal dynamics. There are a couple of remarkable facts associated with that. First, we needed only to look at stationary solutions along the diffusion flow. Second, contrary to standard general relativity these profiles correspond to exact dynamics even when the intrinsic curvature $\textsc{k}$ is negative definite. In particular, de Sitter is an exact solution for a nonconstant scalar field profile, also in an open universe. When the nonlocal operators are chosen to be conformal, for models with $f(R)=-\a_* R$ we have found the general solution for \emph{any} flat or open FRW background, embodied by Eq.~\Eq{geso}; these results have been extended to the inclusion of a barotropic fluid, Eq.~\Eq{vptau12}, also for stringlike actions. Within this class there are solutions without big bang singularity, but there also exist an infinite number of solutions with big bang. Therefore we incline not to link nonsingular solutions with the ultraviolet structure of the nonlocal action. 

At any rate, the space of solutions is likely to be much larger than the portion we have explored here. All our exact solutions have a quadratic potential. Highly nonlinear equations of motion are of great interest, especially in string theory, but the exact solutions can give some indication of the behaviour for general potentials near a local minimum. Cosmological friction modifies the dynamics of nonlocal scalars with respect to Minkowski and, in particular, should drastically change the rolling of the tachyon in string field theory.





\appendix*
\section{de Sitter and power-law solutions in standard local cosmology}

Consider a $D$-dimensional universe filled only with a scalar field with potential $V$ and $\Lambda=0$. The standard Friedmann and continuity equations are
\ba
&&\left(\frac{D}2-1\right)H^2=\frac{\k^2}{D-1}\left(\frac{\dot\phi^2}{2}+V\right)-\frac{\textsc{k}}{a^2}\,,\\
&&H^2+\dot H=\frac{\k^2}{D-1}\left[\frac{2}{D-2}V-\dot\phi^2\right]\,,\\
&&0=\ddot\phi+(D-1)H\dot\phi+V'\,.
\ea
In de Sitter, Eq.~\Eq{dS1}, for a flat universe the exact solution is just a cosmological constant,
\be
\phi(t)=\phi_0\,,\qquad V(\phi)=\frac{(D-1)(D-2)H^2}{2\k^2}\,,\qquad \textsc{k}=0\,.
\ee
The Friedmann equations show that there is no solution if $\textsc{k}=-1$, while there is one for a closed universe, but only in $D=4$:
\be\nonumber
\phi_\pm(t) =\pm \sqrt{\frac{2}{\k^2 H^2}}\rme^{-H t}\,,\qquad \textsc{k}=1\,.
\ee
This solution is \emph{not} de Sitter because spatial sections are not flat. (We recall that ``de Sitter'' in a cosmological sense, $H=\mbox{const}$, corresponds to the mathematical de Sitter spacetime only if spatial sections are flat. In that case, the de Sitter hyperboloid is only half covered by FRW coordinates.) The continuity equation fixes the potential:
\be
V(\phi)=\frac{3H^2}{\k^2}+H^2\phi^2\,.
\ee
The scalar field $\phi_\pm$ rolls down its potential from $t=-\infty$ and climbs it again after passing the global minimum. The solution is actually unique, since cosmological equations of motion are invariant under time reversal, and it does not matter the direction of the rolling in a symmetric potential.

For a power-law expansion,
\be
a(t)=t^p\,,\qquad H(t)=\frac{p}{t}\,,
\ee
one can try the profile $\phi(t)=(\phi_0/q) t^q$ in the (sum of the) Friedmann equations, but one soon finds that it must be $q=0$. This suggests to consider the limit $q\to 0$, which is a logarithmic profile:
\be
\phi(t)=\phi_0\ln t\,.
\ee
This gives
\be
V(\phi)=\frac{(D-1)p-1}{2}\phi_0^2\,\rme^{-2\phi/\phi_0}\,.
\ee
If the universe is flat,
\be
\phi_0=\pm\sqrt{\frac{(D-2)p}{\k^2}}\,,\qquad \textsc{k}=0\,,
\ee
while for a curved universe the case $p=1$ is the only solution:
\be
\phi_0=\pm\sqrt{\frac{D-2+2\textsc{k}}{\k^2}}\,,\qquad p=1\,.
\ee
This solution is real if $D>2(1-\textsc{k})$. Therefore, it is always valid for a closed universe, while for an open universe it exists only in $D>4$.

In the great majority of applications in the literature, the curvature is ignored because its contribution is washed away by inflation. However, it is interesting to note that the only curved solution in ``de Sitter'' is a closed four-dimensional universe, while for a power-law expansion both signs of the curvature are allowed but for $D\geq 5$. In $D=4$, only the closed solution is allowed.

The nonlocal cosmologies described in the main body of the paper, on the other hand, only allow flat and open solutions, but without constraints on the dimensionality of spacetime. In this sense, nonlocal cosmologies are ``complementary'' to the usual ones.


\end{document}